\documentclass[aps,prb,reprint,amsmath,amssymb,superscriptaddress,showpacs]{revtex4-2}
\usepackage{graphicx}
\usepackage{dcolumn}
\usepackage{bm}
\usepackage[colorlinks,linkcolor=blue,anchorcolor=blue,citecolor=blue,urlcolor=blue,filecolor=blue,menucolor=blue,runcolor=blue]{hyperref}
\usepackage{titlesec}
\titleformat{\section}[block]{\normalfont\large\bfseries}{\thesection}{1em}{}
\titleformat{\subsection}[block]{\normalfont\normalsize\bfseries}{\thesubsection}{1em}{}
\renewcommand\thesection{\arabic{section}}
\renewcommand{\thesubsection}{\thesection.\arabic{subsection}}

\begin{document}
\title{$3d$ flat bands and coupled $4f$ moments in the kagome-honeycomb permanent magnet Sm$_{\rm 2}$Co$_{\rm 17}$}
\author{Hao Zheng}
\affiliation{School of Physics, Zhejiang University, Hangzhou 310058, China}
\affiliation{Center for Correlated Matter, Zhejiang University, Hangzhou 310058, China}
\author{Zhiguang Xiao}
\affiliation{School of Physics, Zhejiang University, Hangzhou 310058, China}
\affiliation{Center for Correlated Matter, Zhejiang University, Hangzhou 310058, China}
\author{Ze Pan}
\affiliation{School of Physics, Zhejiang University, Hangzhou 310058, China}
\affiliation{Center for Correlated Matter, Zhejiang University, Hangzhou 310058, China}
\author{Guowei Yang}
\affiliation{School of Physics, Zhejiang University, Hangzhou 310058, China}
\affiliation{Center for Correlated Matter, Zhejiang University, Hangzhou 310058, China}
\author{Yonghao Liu}
\affiliation{School of Physics, Zhejiang University, Hangzhou 310058, China}
\author{Jianzhou Bian}
\affiliation{School of Physics, Zhejiang University, Hangzhou 310058, China}
\author{Yi Wu}
\affiliation{School of Physics, Zhejiang University, Hangzhou 310058, China}
\affiliation{Center for Correlated Matter, Zhejiang University, Hangzhou 310058, China}
\author{Teng Hua}
\affiliation{School of Physics, Zhejiang University, Hangzhou 310058, China}
\affiliation{Center for Correlated Matter, Zhejiang University, Hangzhou 310058, China}
\author{Jiawen Zhang}
\affiliation{School of Physics, Zhejiang University, Hangzhou 310058, China}
\affiliation{Center for Correlated Matter, Zhejiang University, Hangzhou 310058, China}
\author{Jiayi Lu}
\affiliation{School of Physics, Zhejiang University, Hangzhou 310058, China}
\author{Jiong Li}
\affiliation{Shanghai Synchrotron Radiation Facility, Shanghai Advanced Research Institute, Chinese Academy of Sciences, Shanghai 201204, China}
\author{Tulai Sun}
\affiliation{Center for Electron Microscopy, Zhejiang University of Technology, Hanghzhou 310014, China}
\author{Yu Song}
\affiliation{School of Physics, Zhejiang University, Hangzhou 310058, China}
\affiliation{Center for Correlated Matter, Zhejiang University, Hangzhou 310058, China}
\author{Ruihua He}
\affiliation{Key Laboratory for Quantum Materials of Zhejiang Province, Department of Physics, and Zhejiang Institute for Advanced Light Source, Westlake University, Hangzhou 310058, China}
\author{J. Larrea Jim\'enez}
\affiliation{Laboratory for Quantum Matter under Extreme Conditions, Instituto de F\'isica, Universidade de S\~{a}o Paulo, 05508-090 São Paulo, S\~{a}o Paulo, Brazil}
\author{Guanghan Cao}
\affiliation{School of Physics, Zhejiang University, Hangzhou 310058, China}
\affiliation{Interdisciplinary Center for Quantum Information, and State Key Laboratory of Silicon and Advanced Semiconductor Materials, Zhejiang University, Hangzhou 310058, China}
\affiliation{Collaborative Innovation Center of Advanced Microstructures, Nanjing University, Nanjing 210093, China}
\author{Huiqiu Yuan}
\affiliation{School of Physics, Zhejiang University, Hangzhou 310058, China}
\affiliation{Center for Correlated Matter, Zhejiang University, Hangzhou 310058, China}
\affiliation{Collaborative Innovation Center of Advanced Microstructures, Nanjing University, Nanjing 210093, China}
\author{Yuanfeng Xu}
\affiliation{School of Physics, Zhejiang University, Hangzhou 310058, China}
\affiliation{Center for Correlated Matter, Zhejiang University, Hangzhou 310058, China}
\author{Yi Yin}
\affiliation{School of Physics, Zhejiang University, Hangzhou 310058, China}
\affiliation{Collaborative Innovation Center of Advanced Microstructures, Nanjing University, Nanjing 210093, China}
\author{Ming Shi}
\email {shi20001231@zju.edu.cn}
\affiliation{School of Physics, Zhejiang University, Hangzhou 310058, China}
\affiliation{Center for Correlated Matter, Zhejiang University, Hangzhou 310058, China}
\author{Chao Cao}
\email {ccao@zju.edu.cn}
\affiliation{School of Physics, Zhejiang University, Hangzhou 310058, China}
\affiliation{Center for Correlated Matter, Zhejiang University, Hangzhou 310058, China}
\author{Yang Liu}
\email {yangliuphys@zju.edu.cn}
\affiliation{School of Physics, Zhejiang University, Hangzhou 310058, China}
\affiliation{Center for Correlated Matter, Zhejiang University, Hangzhou 310058, China}
\affiliation{Collaborative Innovation Center of Advanced Microstructures, Nanjing University, Nanjing 210093, China}
\date{\today}%

\addcontentsline{toc}{chapter}{Abstract}

\begin{abstract}
Rare earth permanent magnets (REPMs) with both localized moments and itinerant conduction bands are not only important for fundamental research but also have significant technological applications. In particular, Sm$_{\rm 2}$Co$_{\rm 17}$ is a prototypical high-temperture REPM, where the Co atoms form a kagome-honeycomb stacked lattice. Here we report synthesis of epitaxial Sm$_{\rm 2}$Co$_{\rm 17}$ films using molecular beam epitaxy and measurements of their momentum-resolved electronic structure from \textit{in-situ} angle-resolved photoemission spectroscopy. Our results unveil two flat bands from Co $3d$ orbitals near the Fermi level ($E_F$), one at $\sim$\,--300\,meV and another right at $E_F$, which arise from orbital-selective destructive interference and strong electron correlations, respectively. In addition, our results reveal that Sm $4f$ states are far away from $E_F$ (hence mostly localized) and exhibit an anomalous temperature dependence, caused by the $3d$-$4f$ magnetic coupling. 
Our findings provide direct spectroscopic insights to understand the strong uniaxial ferromagnetism in Sm$_{\rm 2}$Co$_{\rm 17}$ (and REPMs alike). Our work also opens avenues to explore flat-band physics near $E_F$ and emergent phenomena in correlated kagome-honeycomb lattices.

\vspace{10pt}
\noindent \textbf{kagome lattice, flat bands, rare-earth permanent magnet, molecular beam epitaxy, angle-resolved photoemission spectroscopy}

\vspace{10pt}
\noindent \textbf{PACS number(s): }71.20.Be, 71.20.Eh, 73.22.-f, 75.50.Ww, 81.15.-z
\end{abstract}

\maketitle

\section{Introduction}
Rare-earth permanent magnets (REPMs), which are intermetallic compounds composed of rare earth ($R$) and transition metal ($T$) elements, are extensively utilized in the automotive, data storage, electric power, and medical imaging industries. The key feature driving their widespread application is their strong uniaxial ferromagnetism (FM), which remains robust against demagnetization. Significant efforts have been dedicated to elucidating the fundamental mechanisms underpinning the robust FM in REPMs \cite{RFeBRMP1991,RxTyreviewRPP1977,CoeyIEEE2011,AMreview2011,REPMreview2018,fangSCPMA2025,liuSCPMA2025}, aiming to design and engineer superior REPM materials. While the strong uniaxial FM is typically explained by the magnetic exchange interactions of the $3d$ and $4f$ sublattices \cite{campbell1972,MolecularFieldInteraction-JAP1991}, particularly the $3d$-$3d$ and $3d$-$4f$ interactions, the different types of exchange interactions are expected due to the localized or itinerant nature of $4f$/$3d$ electrons. Therefore, it is crucial to reveal their energy positions relative to Fermi level ($E_F$) and momentum dispersions for a fundamental understanding of REPMs. However, direct measurements of their momentum-resolved electronic structures remain elusive up to now.

Among REPMs, Sm$_{\rm 2}$Co$_{\rm 17}$ stands out due to their exceptionally strong FM and very high Curie temperatures ($T_c$) \cite{RxTyreviewRPP1977}. Interestingly, it belongs to a large family of binary compounds R$_{\rm x}$T$_{\rm y}$ ($x$:$y$ being 1:5, 2:17, 1:2, etc.), which feature kagome layers composed of transition metal atoms $T$ (see Figure~\ref{Fig1}(a-c)). Recently, kagome-based materials have generated widespread interest in physics and material science \cite{nature2022,NRP2023}, owing to a variety of many-body phases including superconductivity \cite{XGWPRB2009,JXL2012PRB,Ortiz2020PRL}, FM \cite{mielke1991-1,FBFMreview1998,Yenature2018,Yinnature2018}, spin/charge density waves \cite{ThomalePRL2013,QHW2013PRB,Nienature2022,Tengnature2022}, Wigner crystal \cite{WignerCreystal}, and fractional quantum Hall effect \cite{FQHE1,FQHE2,FQHE3}. In particular, destructive interference of electronic wave functions in the kagome lattice can give rise to flat bands with a high density of states (DOS), as recently observed by angle-resolved photoemission spectroscopy (ARPES) \cite{Fe3Sn2-PRL2018,CoSn-NC2020liu,CoSn-NC2020kang,FeSn-NM2020,Fe3Sn2nature2024,CoSn-PRL2022,XV6Sn6-NP2023,RbTi3Bi5-NP2023, zhengSCPMA2024}.
However, in most cases, the flat bands are far away from $E_F$ and they do not make significant contributions to the physical properties. If the flat bands are close to the $E_F$, they can be favorable for the formation of flat-band FM \cite{mielke1991-1,FBFMreview1998}. In Sm$_{\rm 2}$Co$_{\rm 17}$, the kagome Co layers are sandwiched by honeycomb Co layers, and symmetry-protected flat bands were predicted in such a kagome-honeycomb lattice \cite{FBnature2022,FBNP2022,FB-PRB2015}, motivating the spectroscopic detection of these flat bands and the exploration of their connection to the robust FM.

A typical method for measuring the momentum-dependent electronic structure is ARPES. However, this is challenging for REPMs, as ARPES measurements typically require large single crystals with natural cleavage planes and minimal residual magnetic field. For Sm$_{\rm 2}$Co$_{\rm 17}$, single crystals are very difficult to synthesize \cite{2019AIP}, hampering in-depth understanding of its intrinsic properties. Here, we successfully overcome these experimental obstacles by combining thin film growth via molecular beam epitaxy (MBE) with \textit{in-situ} ARPES measurements \cite{Monkman2012NM}, thereby obtaining for the first time the momentum-resolved electronic structure of a prototypical REPM Sm$_{\rm 2}$Co$_{\rm 17}$. Our work paves the way for a fundamental understanding of the strong uniaxial ferromagnetism in Sm$_{\rm 2}$Co$_{\rm 17}$ (and other REPMs) from electronic structure perspective and represents a step forward to search for flat bands beyond the simple kagome lattice \cite{FBnature2022,FBNP2022}. Our synthesis of epitaxial Sm$_{\rm 2}$Co$_{\rm 17}$ films also opens up opportunites to explore thin-film applications based on REPMs.

\section{Experimental and calculational methods}

The epitaxial Sm$_{\rm 2}$Co$_{\rm 17}$ films were grown on Al$_{\rm 2}$O$_{\rm 3}$(0001) substrates, modified from a previous growth method \cite{SmCoACS2021}. The base pressure of the MBE chamber is better than $1\times10^{-10}$ mbar. The evaporations of Sm and Co were achieved using effusion cells, whose deposition rates were set to approximately $\sim$0.8 {\AA}~min$^{-1}$ for Sm and $\sim$1.4 {\AA}~min$^{-1}$ for Co, respectively, as determined by a quartz crystal monitor. Before growth, Al$_{\rm 2}$O$_{\rm 3}$ substrates were heated up to 650 ℃ in ultrahigh vacuum for 30 minutes. We found it helpful to first grow a thin Co buffer layer at 120 ℃ before the actual growth of Sm$_{\rm 2}$Co$_{\rm 17}$. The substrates were then heated to 400 ℃ for growing Sm$_{\rm 2}$Co$_{\rm 17}$ films. The growth of Sm$_{\rm 2}$Co$_{\rm 17}$ films were monitored by \textit{in-situ} reflection high-energy electron diffraction (RHEED). After growth, the Sm$_{\rm 2}$Co$_{\rm 17}$ films (typically 20 nm) were transferred under ultrahigh vacuum to a lab-based ARPES system equiped with a helium lamp for \textit{in-situ} ARPES measurements. The sample temperature was kept at $\sim6$ K during all measurements unless noted otherwise. All ARPES data were taken with the He I$\alpha$ (21.2 eV) or He II (40.8 eV) photons. The typical energy and momentum resolution is $\sim$12 meV and $\sim$0.01 {\AA}$^{-1}$, respectively.

Density Functional Theory (DFT) calculations were performed using the Vienna \emph{Ab initio} Simulation Package (VASP) \cite{VASP1993,PAW1999}, incorporating the Perdew--Burke--Ernzerhof (PBE) generalized gradient approximation (GGA) \cite{PBE1996} for the exchange-correlation functional. To model the ferromagnetic (FM) state, Sm $4f$ electrons were treated as valence electrons with a Hubbard interaction $U=7.0$\,eV and Hund's coupling $J=1.0$\,eV, as is commonly adopted for Sm-based systems. A linear interpolation double-counting (DC) scheme between fully localized limit (FLL) DC and around-mean-field (AMF) DC \cite{PhysRevB.67.153106,YbPtBiPRB2023} which minimize the total energy was chosen to give a resonable magnetic moment of Sm. For Co $3d$ electrons, we found that $U=J=0$ gave the best agreement with our experimental results (see Supporting Information for details). To achieve a quantitative match with experiment, we applied a bandwidth renormalization (reduction) factor of $\sim 1.7$, which is not uncommon in strongly correlated $3d$-electron systems as DFT calculations tend to overestimate the bandwidth \cite{King20142014NN,PhysRevB.89.220506,CoSnnanoletter2023,PhysRevB.109.104410}. 

The structure of Sm$_{\rm 2}$Co$_{\rm 17}$ films was examined using \textit{ex-situ} high-angle annular dark-field scanning transmission electron microscopy (HAADF-STEM) and X-ray diffraction (XRD). HAADF-STEM imaging was carried out on a state-of-the-art spherical-aberration-corrected transmission electron microscope. Synchrotron XRD and X-ray absorption spectroscopy (XAS) were performed at beamlines BL02U2, BL07U, and BL11B of the Shanghai Synchrotron Radiation Facility (SSRF). Magnetic properties were studied with a SQUID magnetometer at low temperatures, subtracting the substrate signal. To minimize contributions from the Co buffer, these \textit{ex-situ} measurements were done on thicker Sm$_{\rm 2}$Co$_{\rm 17}$ layers (typically 100\,nm) with a very thin or no Co buffer.

\section{Results}

\subsection{Characterization of Sm$_{\rm 2}$Co$_{\rm 17}$ films}

\begin{figure*}[ht]
  \centering
  \includegraphics[width=0.7\linewidth]{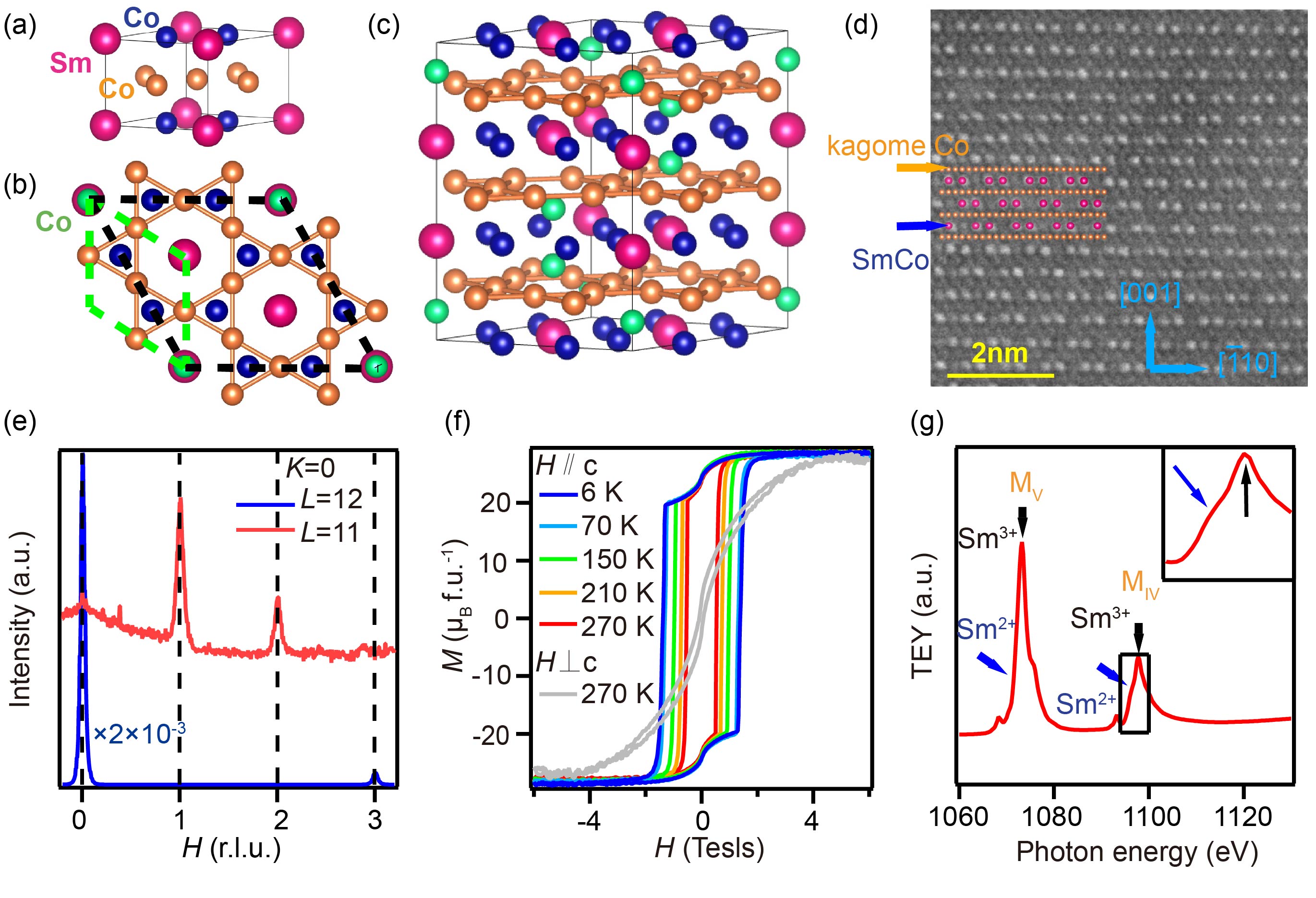}
  \caption{
  (a) The primitive cell of SmCo$_{\rm 5}$. 
  (b) Top view of the unit cell of Sm$_{\rm 2}$Co$_{\rm 17}$. 
  (c) The conventional cell of rhombohedral Sm$_{\rm 2}$Co$_{\rm 17}$. The yellow and blue balls in (a-c) indicate Co atoms in the kagome and honeycomb layers, respectively. 
  (d) HAADF-STEM image of a Sm$_{\rm 2}$Co$_{\rm 17}$ film. Yellow and blue arrows indicate the kagome Co and SmCo layers. 
  (e) Synchrotron XRD scans along $H$ at two L values (note the different multiplication factors). 
  (f) The magnetization $M$ as a function of field $H$ at different temperatures for fields parallel and perpendicular to $c$ axis.
  (g) XAS result of a thick Sm$_{\rm 2}$Co$_{\rm 17}$ film near the Sm $M$ edge, obtained from total electron yield (TEY). The inset shows a zoom-in view near the $M_{IV}$ edge, highlighting a very weak Sm$^{\rm 2+}$ component. The Sm$^{\rm 3+}$ and Sm$^{\rm 2+}$ components are marked by black and blue arrows, respectively.  
  }
  \label{Fig1}
\end{figure*}

The fundamental building block of R$_{\rm x}$T$_{\rm y}$ compounds is the hexagonal RT$_{\rm 5}$ structure, as illustrated in Figure~\ref{Fig1}(a) (taking SmCo$_{\rm 5}$ as an example). This structure consists of alternating SmCo$_2$ and Co layers stacked along the $c$ axis. Within the Co layer, the Co atoms form a pure kagome lattice without interstitial atoms, while in the SmCo$_2$ layer, Sm and Co atoms are arranged in triangular and honeycomb lattices, respectively. By periodically replacing one-third of the Sm atoms in the SmCo$_2$ layer with a Co-Co dimer (green balls in Figure~\ref{Fig1}(b,c)), the rhombohedral phase of Sm$_{\rm 2}$Co$_{\rm 17}$ can be obtained. Hence, the Sm$_{\rm 2}$Co$_{\rm 17}$ lattice can be viewed as a weakly distorted version of the basic SmCo$_{\rm 5}$ lattice, with an enlarged unit cell of $a^* \approx \sqrt{3}\,a$ and $c^* \approx 3\,c$, where $a^*$ and $c^*$ ($a$ and $c$) are the lattice constants of Sm$_{\rm 2}$Co$_{\rm 17}$ (SmCo$_{\rm 5}$). Notably, the kagome and honeycomb Co layers remain intact in the Sm$_{\rm 2}$Co$_{\rm 17}$ structure (yellow and blue balls in Figure~\ref{Fig1}(b,c)).

We successfully grow high-quality Sm$_{\rm 2}$Co$_{\rm 17}$ films using MBE (See Figure~S1 in Supporting Information (SI) for more details). The sample quality is verified by the atomic-resolution images obtained via HAADF-STEM (Figure~\ref{Fig1}(d)) and the sharp peaks from synchrotron XRD (Figure~\ref{Fig1}(e)). 
Specifically, since one third of Sm atoms in Sm$_{\rm 2}$Co$_{\rm 17}$ is replaced periodically with a Co-Co dimer compared to SmCo$_{\rm 5}$, the Sm atoms in the SmCo layer exhibits a $\times$3 periodicity along $[\bar{1} 1 0]$, which can be clearly identified in Figure~\ref{Fig1}(d) as two bright spots and one weak spot. The XRD result in Figure~\ref{Fig1}(e), where the $H$, $K$, and $L$ directions are defined with respect to the bulk Sm$_{\rm 2}$Co$_{\rm 17}$ lattice, shows that the superstructure peaks exclusively associated with Sm$_{\rm 2}$Co$_{\rm 17}$, e.g., ($H$,$K$,$L$)=(1,0,11), can be observed, although they are nearly three orders of magnitude weaker than the dominant peaks expected from ideal SmCo$_{\rm 5}$, e.g., ($H$,$K$,$L$)=(0,0,12). This indicates that the superstructure distortion from SmCo$_{\rm 5}$ to Sm$_{\rm 2}$Co$_{\rm 17}$ is weak but clearly discernible.

The magnetization as a function of applied field ($M$-$H$ curves) is summarized in Figure~\ref{Fig1}(f), clearly showing ferromagnetic behavior with pronounced hysteresis loops. When the field is aligned along the $c$~axis (the easy axis), the saturation moment remains nearly unchanged up to 270\,K, consistent with a high Curie temperature ($T_C$). In contrast, the coercive field ($H_c$) is strongly temperature-dependent (Figure~\ref{Fig1}(f) and Figure~S1(d) in SM): for instance, $H_{c}$ increases from $\sim$0.5\,T at 270\,K to $\sim$1.4\,T at 6\,K. The strong magnetocrystalline anisotropy (MCA), which is crucial for rare-earth permanent magnet (REPM) applications, is evident from the $M$-$H$ response measured with the field perpendicular to the $c$~axis (Figure~\ref{Fig1}(f)), where the moment remains unsaturated even under a field of $\sim$4\,T and the hysteresis is significantly reduced. Therefore, our magnetic measurements of single-crystal thin films confirm the strong MCA expected in Sm$_{\rm 2}$Co$_{\rm 17}$ and also reveal its pronounced temperature dependence, which can be related to the anomalous temperature evolution of $4f$ states (Figure~\ref{Fig5}(a,b)). Finally, measurements from XAS (Figure~\ref{Fig1}(g)) show dominant contributions from Sm$^{\rm 3+}$ (strong peaks marked by black arrows), and very weak Sm$^{\rm 2+}$ components (shoulder structures marked by blue arrows).

\begin{figure}[ht]
  \centering
  \includegraphics[width=1\linewidth]{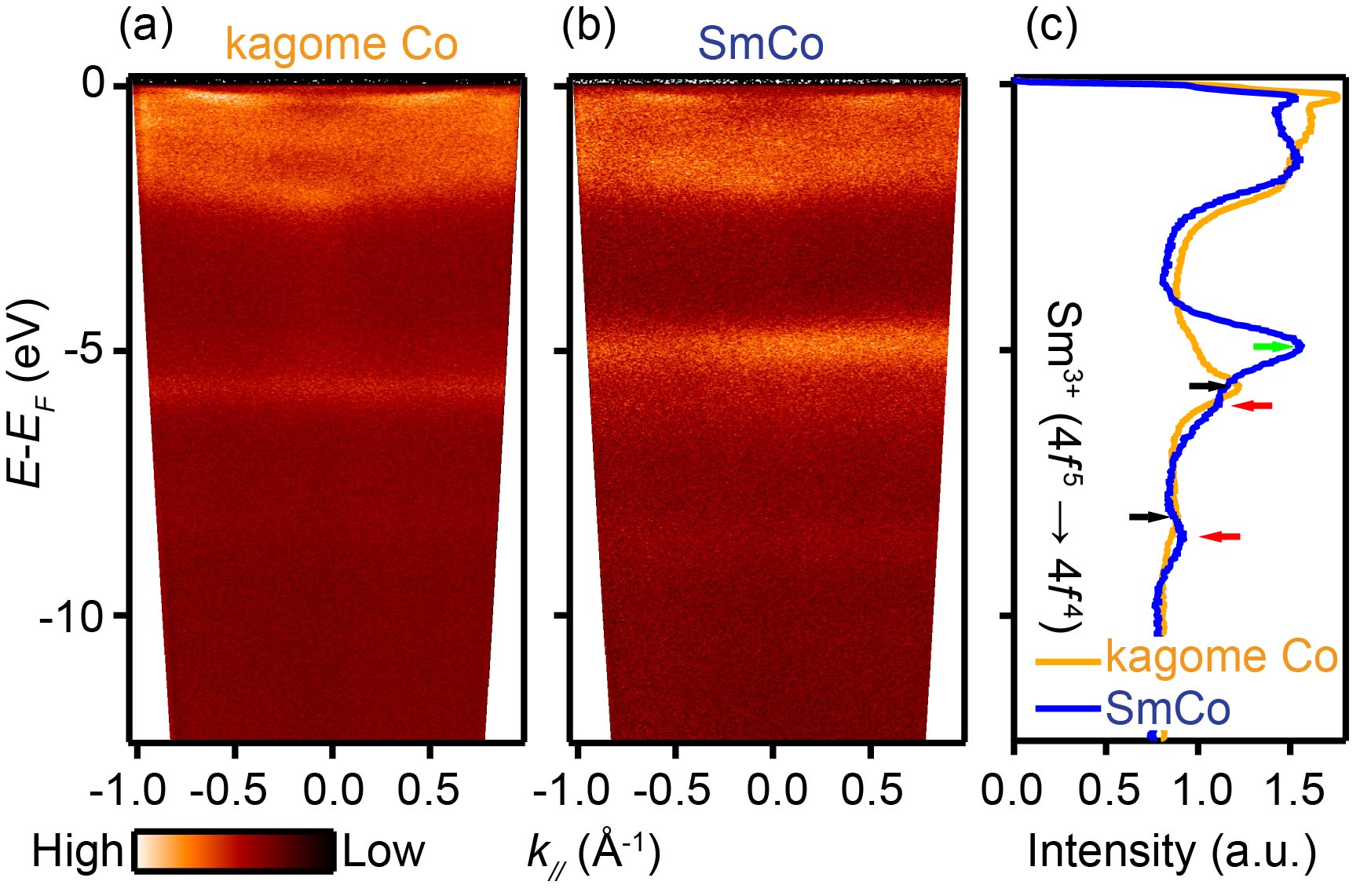}
  \caption{
    (a,b) Large-scale ARPES spectra of kagome Co terminated surface (a) and SmCo terminated surface (b) using 40.8 eV photons. 
    (c) The corresponding momentum-integrated EDCs of two terminations. The black and red arrows indicate the bulk and surface-shifted $4f$ peaks, while the green arrow indicates a surface $4f$ peak on the SmCo termination. 
  }
  \label{Fig2}
\end{figure}

An advantage of MBE growth is the ability to control the surface termination. Figure~\ref{Fig2} compares the energy distribution curves (EDCs) from photoemission measurements for kagome Co and SmCo terminated surfaces. For the kagome Co termination, two flat bands at $-5.6\,\mathrm{eV}$ and $-8\,\mathrm{eV}$ (indicated by black arrows in Figure~\ref{Fig2}(c)) can be clearly observed. These features can be attributed to bulk $4f$ states associated with Sm$^{3+}$ ($4f^5$), consistent with previous studies on Sm-based compounds \cite{SmCo2-PRB1993,SmB6ARPESreview2014,SmRh2Si2-PRB2017}. In contrast, for the SmCo termination, these $4f$ states shift downward by approximately $0.4\,\mathrm{eV}$ (red arrows), likely due to a surface core-level shift \cite{SmCo2-PRB1993,PhysRevB.51.7920,JPCL13.7861}, while an intense peak emerges near $-5\,\mathrm{eV}$ (green arrow), possibly a pure surface component \cite{JPCL13.7861,PhysRevB.106.155136}. Moreover, the valence bands from $-2\,\mathrm{eV}$ to $E_F$ are obviously stronger for the kagome Co termination, as they predominantly originate from Co $3d$ electrons (see below).

\subsection{Fermi surface and flat bands on kagome Co termination}

\begin{figure*}[ht]
  \includegraphics[width=\linewidth]{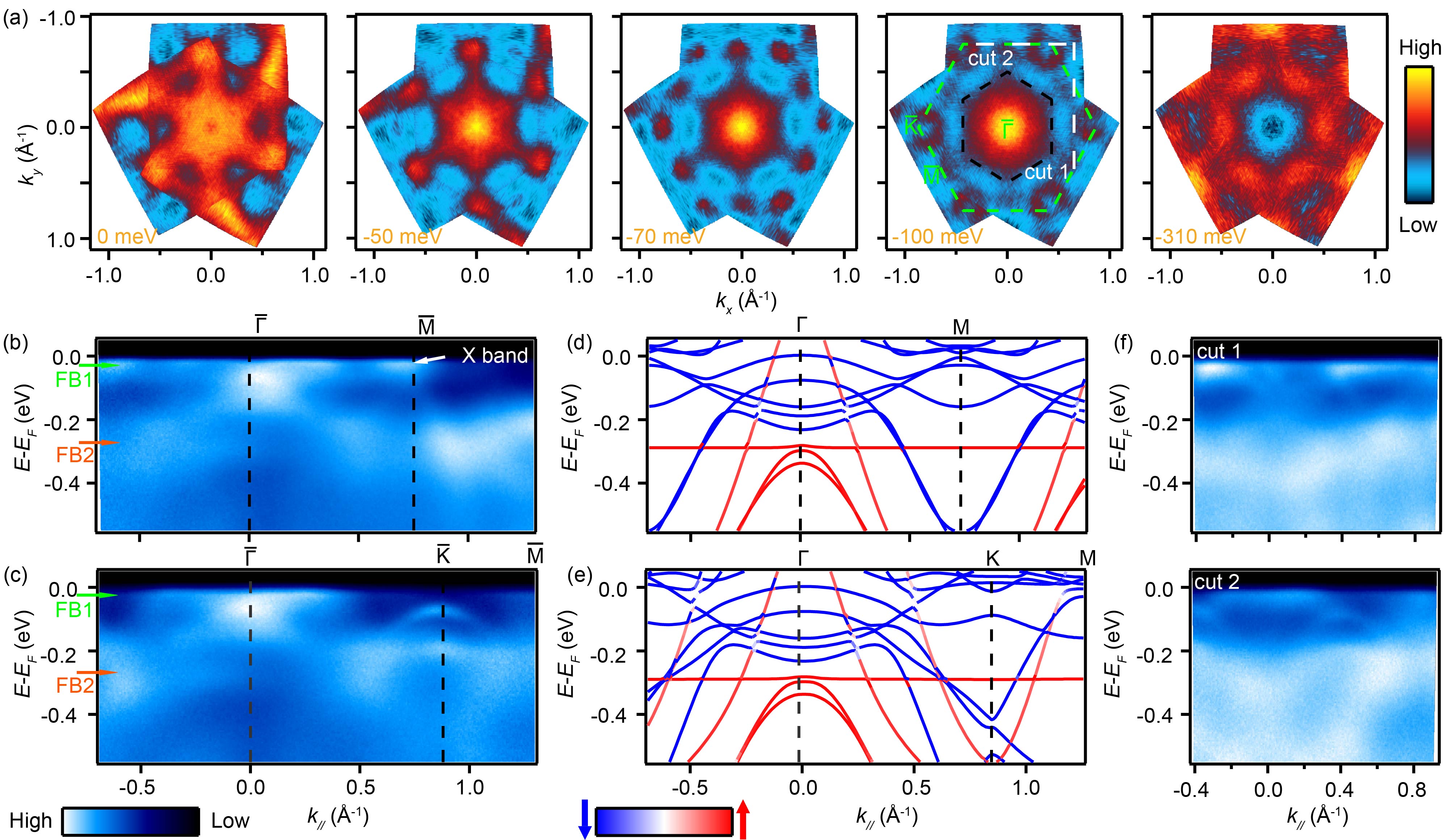}
  \centering
  \caption{Electronic structure for kagome Co terminated surface.
  (a) Constant energy $k_{x}$-$k_{y}$ maps at various energies. The original data, covering more than half of the surface BZ, were symmetrized according to the bulk $C_{3}$ symmetry to yield these maps. Green and black dashed hexagons on the -100 meV map indicate the surface BZ's of SmCo$_{\rm 5}$ and Sm$_{\rm 2}$Co$_{\rm 17}$, respectively.
  (b,c) Experimental energy-momentum cuts along $\bar{\Gamma}$-$\bar{M}$ (b) and $\bar{\Gamma}$-$\bar{K}$-$\bar{M}$ (c).
  (d,e) Spin-resolved band structure from DFT+$U$ along $\Gamma-M$ (d) and $\Gamma-K-M$ (e), for comparison with (b,c). The red and blue curves indicate majority-spin and minority-spin bands, respectively.
  (f) Experimental energy-momentum cuts along two directions labelled in the -100 meV map in (a).
  }
  \label{Fig3}
\end{figure*}

Figure~\ref{Fig3}(a) presents constant-energy $k_x$-$k_y$ maps on the kagome Co terminated surface, acquired from ARPES using 21.2\,eV photons. This corresponds to $k_z \sim 4\pi/c$ based on an estimated inner potential $V_0=19.6$\,eV, determined by matching the ARPES data taken with both 21.2\,eV and 40.8\,eV photons to the calculations (see Figure~S2 in SI for data taken by 40.8\,eV photons). Notably, the dominant features in these maps follow the surface Brillouin zone (BZ) of the Co kagome layer, which is the same as SmCo$_{\rm 5}$ (green dashed hexagon at $E=-100$\,meV), rather than the smaller BZ of Sm$_{\rm 2}$Co$_{\rm 17}$ (black dashed hexagon). For instance, at $E=-100$\,meV, circular hole pockets are clearly observed at the $\bar{K}$ points of the SmCo$_{\rm 5}$ BZ. This hole band at $\bar{K}$ is evident in the energy-momentum cut of Figure~\ref{Fig3}(c), where the band maximum is located near $-70$\,meV. Meanwhile, an electron pocket centered at $\bar{M}$ emerges around $-70$\,meV (Figure~\ref{Fig3}(a)) and expands upon approaching $E_F$, corresponding to the upper branch of an $X$-shaped band at $\bar{M}$ (white arrow in Figure~\ref{Fig3}(b)). The fact that the measured bands follow the surface BZ of the Co kagome layer (or SmCo$_{\rm 5}$) indicates that the surface-sensitive ARPES spectra are mostly contributed by electronic states residing on the Co kagome layer and are therefore not sensitive to the weak lattice distortion of the SmCo$_{\rm 2}$ layer in Sm$_{\rm 2}$Co$_{\rm 17}$. Hence, for simplicity, we compare our ARPES data with the simpler band structure of SmCo$_{\rm 5}$, as shown in Figure~\ref{Fig3}(d,e). Indeed, most of the spectral features can be reasonably captured by this simplified calculation. By contrast, many band features predicted from the Sm$_{\rm 2}$Co$_{\rm 17}$ calculation are not observed experimentally (Figure~S3 in SI), further validating our comparison with the simplified band structure of SmCo$_{\rm 5}$ that ignores the weak distortion in the SmCo$_{\rm 2}$ layer.

In Figure~\ref{Fig3}(a), a broad pocket centered at $\bar{\Gamma}$ is observed at $E_F$ ($E=0$), extending over a large momentum range. This feature corresponds to a nearly dispersionless band, labeled FB1 in Figure~\ref{Fig3}(b,c). In addition, another flat band (FB2) appears around $-300$\,meV, exhibiting only slight dispersion (see also the second-derivative data in Figure~S4 in SI for a clearer presentation). Figure~\ref{Fig3}(f) shows two additional energy--momentum cuts, demonstrating that both FB1 and FB2 persist over a wide momentum range.

Figure~\ref{Fig3}(d,e) depict the spin-resolved band structure for ferromagnetic SmCo$_{\rm 5}$, calculated via DFT including Hubbard $U$ and Hund's coupling $J$ (DFT+$U$) (see SI for more details). The calculations indicate that the Sm valence is close to Sm$^{3+}$ ($4f^5$), and correspondingly the $4f$ multiplet lies several eV's below $E_F$, consistent with our XAS and photoemission results (Figure~\ref{Fig1}(g) and Figure~\ref{Fig2}). Moreover, the valence bands near $E_F$ predominantly originate from Co $3d$ electrons and exhibit large spin splitting (Figure~S6 in SI). 
The majority-spin bands (red) are almost fully occupied below $E_F$, whereas the minority-spin bands (blue) dominate near $E_F$. The essential experimental features near $E_F$ are reasonably reproduced (compare Figure~\ref{Fig3}(b,c) and \ref{Fig3}(d,e)). Interestingly, the calculations also reveal a nearly non-dispersive band at $\sim -300$\,meV, which is consistent with the FB2 band observed experimentally.

\begin{figure}[ht]
  \includegraphics[width=1\linewidth]{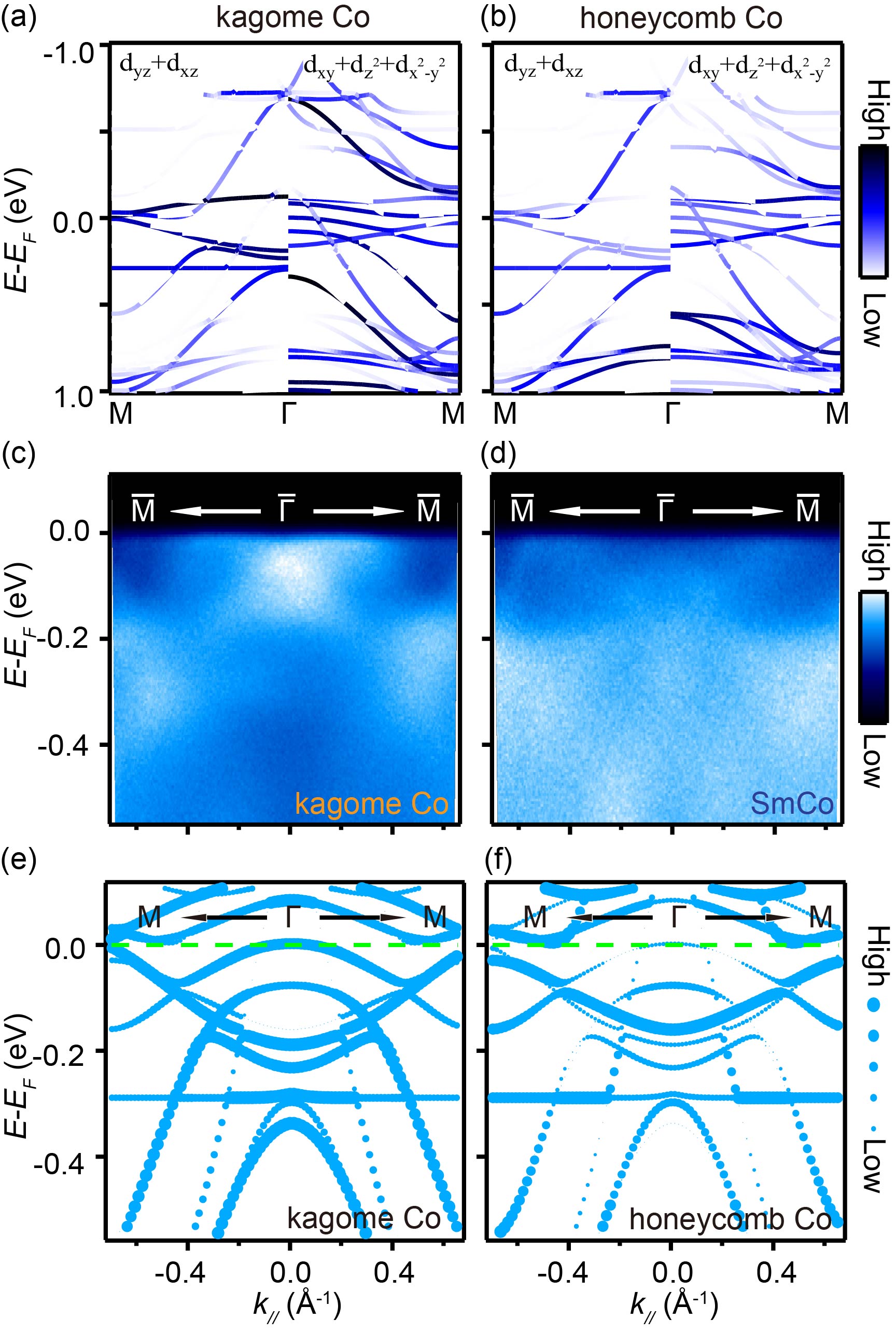}
  \centering
  \caption{
    (a,b) Orbital-resolved DFT+$U$ band structure (left: $d_{xz}$+$d_{yz}$. right: $d_{xy}$+$d_{z^2}$+$d_{x^2+y^2}$) projected from kagome Co (a) and honeycomb Co (b) layers.
    (c,d) ARPES spectra near $E_F$ corresponding to kagome Co (c) and SmCo (d) terminations.
    (e,f) Orbital-integrated layer-resolved band structure from the kagome Co (e) and honeycomb Co (f) layers, for comparison with (c,d). The color bars are shown on the right.
  }
  \label{Fig4}
\end{figure}

The FB1 band at $E_F$ is not adequately captured by the DFT+$U$ calculations.
While it is tempting to attribute this band to Sm $4f$ states, our experimental results are inconsistent with such a scenario: the photoemission intensity of FB1 strongly decreases when using 40.8\,eV photons (Figure~S7 in SI), whereas the photoemission cross section for $4f$ electrons should be enhanced under this photon energy. This indicates that FB1 is predominantly derived from Co $3d$ orbitals rather than Sm $4f$ electrons. This is also consistent with the Sm valence being close to 3+ (see Figure~\ref{Fig5} below), where the Sm $4f$ states near $E_F$ corresponding to Sm$^{2+}$ should be weak. Therefore, the FB1 is most likely caused by the strong correlation effects in a multi-orbital $3d$ system, due to a combination of Hubbard interaction $U$ and Hund's coupling $J$ \cite{Yin2011kinetic,Yu2013PRL}. Such correlation effects can lead to orbital-selective band renormalization, an accurate description of which is beyond the scope of standard DFT methods \cite{Samanta2024NC,CuV2S4NP2024}.

\subsection{Destructive interference in a kagome-honeycomb lattice}

Figure~\ref{Fig4}(a,b) show the orbital- and layer-resolved band structure of SmCo$_{\rm 5}$ obtained from DFT+$U$ calculations (see also Figure~S8 in SI for details). Here the orbitals were grouped into $d_{xz}$+$d_{yz}$ and $d_{xy}$+$d_{x^2+y^2}$+$d_{z^2}$, respectively, based on the wavefunction symmetry and orbital-selective destructive interference discussed in Ref .~\cite{FB-PRB2015}. They demonstrate that the flat band near $-300$\,meV (FB2) originates from Co $d_{xz}$ and $d_{yz}$ orbitals, with contributions from both the kagome and honeycomb layers. By contrast, the minority-spin bands at $E_F$ (corresponding to FB1) arise mainly from $d_{xy}$ and $d_{x^2+y^2}$ orbitals located in the kagome layer. This naturally explains why, on the SmCo terminated surface (Figure~\ref{Fig4}(d)), the FB1 band near $E_F$ is much weaker compared to the kagome Co termination (Figure~\ref{Fig4}(c)), whereas the FB2 band remains strong, thanks to the surface sensitivity of ARPES measurements. Moreover, the layer-resolved (and orbital-integrated) DFT+$U$ calculations in Figure~\ref{Fig4}(e,f) can reasonbaly reproduce the observed spectral features and their termination dependence (comparing Figure~\ref{Fig4}(e,f) with Figure~\ref{Fig4}(c,d)).

The very flat band at $\sim\! -300$\,meV (Figure~\ref{Fig3}(d,e)) and its spatial extension into the honeycomb layer (Figure~\ref{Fig4}(a,b)) are noteworthy, since interlayer hopping and multiorbital mixing often destroy flat bands in kagome lattices. Here, the flat band arises from orbital-selective destructive interference in the kagome-honeycomb stacked lattice, as theoretically proposed in Refs.~\cite{FBnature2022,FB-PRB2015}: the lattice symmetry effectively decouples the $d_{xz}$ and $d_{yz}$ orbitals from other orbitals at $k_z=0$, leading to destructive interference for both intralayer hopping within the kagome layer and interlayer hopping between the kagome and honeycomb layers. This interplay results in a symmetry-enforced flat band at $k_z=0$, namely FB2. Our calculations further indicate that this flat band persists over a considerable range in $k_z$ (Figure~S9 in SI).

\subsection{4$f$ states and temperature evolution}

\begin{figure}[ht]
  \centering
  \includegraphics[width=1\linewidth]{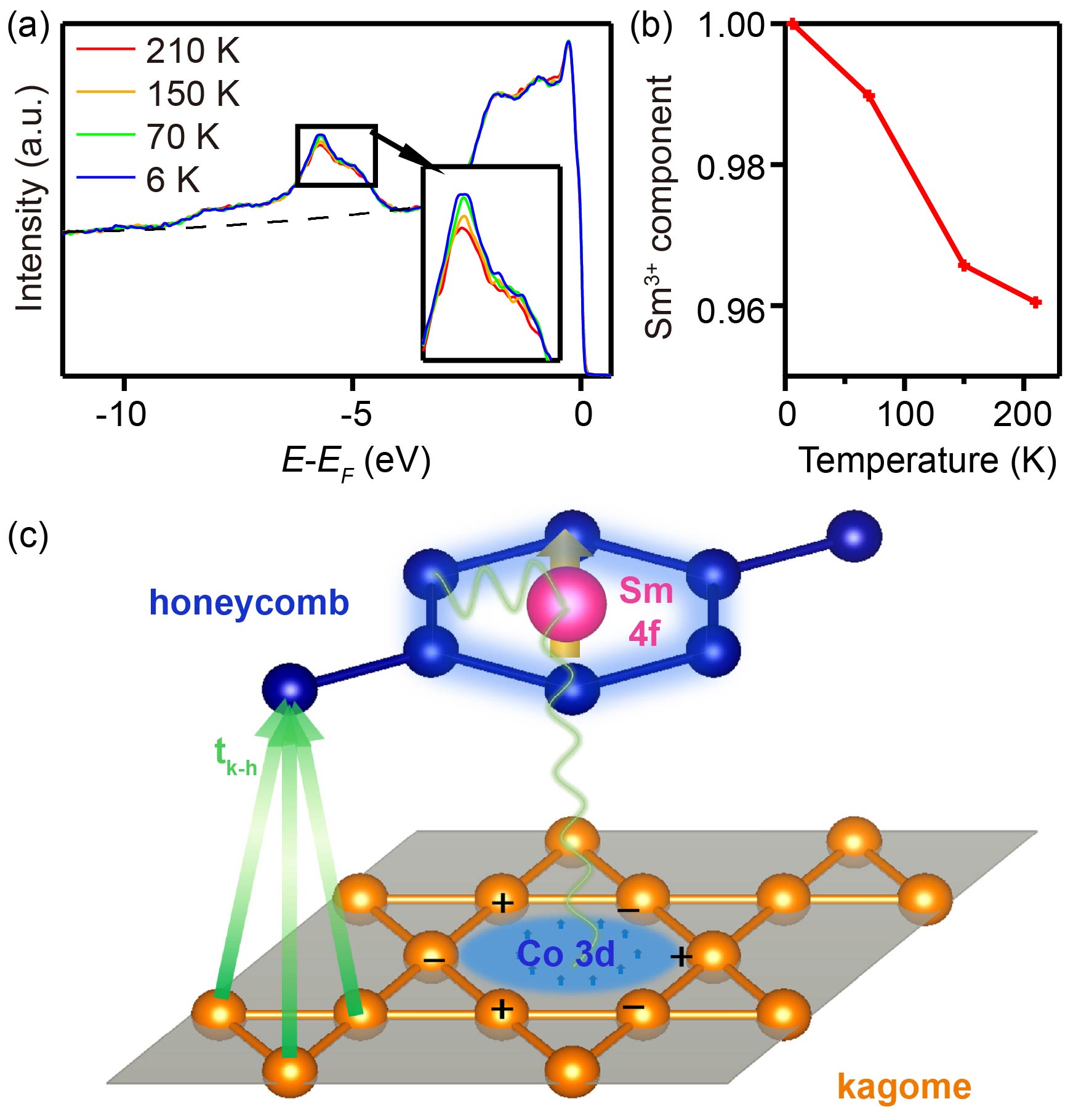}
  \caption{
    (a) Temperature evolution of momentum-integrated EDCs. The inset shows a zoom-in view of the Sm$^{\rm 3+}$ $4f$ states.
    (b) The Sm$^{\rm 3+}$ component extracted from (a) and normalized to its value at 6 K.
    (c) A cartoon illustrating the $3d$-$3d$ and $3d$-$4f$ magnetic interactions. The $3d$-$3d$ interaction stems from itinerant $3d$ bands, which develop flat bands from destructive interferences, for both intralayer hopping within the kagome layer (light blue) and interlayer hopping between the kagome and honeycomb layers (green arrows). The $3d$-$4f$ interaction, indicated by grey wavy lines, is realized through mostly localized $4f$ electrons with suppressed valence fluctuation and contributes to the large MCA.
  }
  \label{Fig5}
\end{figure}

The $4f$ electrons play an important role in generating the large MCA in REPMs \cite{SC5-Tmagmom-JAP1979,3d-4f-book2007,PatrickPRL2018}: the crystal electric field (CEF) experienced by $4f$ electrons produces strongly anisotropic $4f$ moments, which couple to $3d$ moments, leading to large coercive fields and resistance against demagnetization. Typically, $4f$ electrons are regarded as stable, with negligible valence mixing. However, Sm-based materials often exhibit valence fluctuations between Sm$^{3+}$ and Sm$^{2+}$, as exemplified by SmB$_{6}$ \cite{SmB6-XAS-JPCS2009,SmB6NRP2020,SmB6PRL2013}, where the average Sm valence is close to 2.6. In Sm$_{\rm 2}$Co$_{\rm 17}$, both XAS measurements and DFT calculations indicate that the Sm valence remains predominantly $3^+$, albeit with small $2^+$ component. Figure~\ref{Fig5}(a) displays the temperature evolution of the $4f$ states associated with Sm$^{3+}$, whose intensity (proportional to the Sm$^{3+}$ component) gradually increases upon cooling. This phenomenon is corroborated by additional measurements under repeated temperature cycling on different samples (Figure~S11 in SI). 
Moreover, our ARPES data reveal reduction of spectral intensities near $-0.5\,\mathrm{eV}$ and $-1.5\,\mathrm{eV}$ upon cooling, which could indicate either a direct response of the Co $3d$ states due to the $3d$-$4f$ coupling or a reduction in the Sm$^{2+}$ component. Future temperature-dependent resonant photoemission measurements, to be carried out in synchrotron facilities with tunable photon energies, will be crucial for distinguishing these two scenarios.

To quantify the temperature-dependent valence change, we integrated the intensity of the Sm$^{3+}$ $4f$ peak (from $-10\,\mathrm{eV}$ to $-4\,\mathrm{eV}$) after background subtraction and normalized it to its value at 6\,K (Figure~\ref{Fig5}(b)). The analysis indicates a small valence increase of approximately 0.04 from 210\,K to 6\,K, which is difficult to resolve in XAS measurements. Such a low-temperature increase in Sm valence is rather unusual, contrasting with well-known mixed-valent Sm systems \cite{SmB6-XAS-JPCS2009,SmS-DMFT-PRL2015}, where Sm valence typically decreases toward Sm$^{2+}$ at low temperature due to enhanced $4f$ itineracy. Such anomalous temperature evolution could be attributed to the magnetic coupling between Sm $4f$ moments and Co $3d$ bands, which we shall discuss below.

\section{Discussion}

The flat bands observed near $E_F$, arising from destructive interference in the kagome-honeycomb lattice (FB2) and from strong $3d$ correlation effects (FB1), contribute to an enhanced DOS that can promote itinerant FM in Sm$_{\rm 2}$Co$_{\rm 17}$. This mechanism could also apply to other RCo$_{\rm 5}$ and R$_{\rm 2}$Co$_{\rm 17}$ compounds, providing a natural explanation for their common FM behavior \cite{RxTyreviewRPP1977,AMreview2011}. According to calculations \cite{FB-PRB2015,KoudelaPRB2015}, the FB2 band is highly tunable under pressure or strain. Interestingly, experiments on YCo$_{\rm 5}$ (which lacks $4f$ electrons) have revealed a marked magnetoelastic lattice collapse under pressure \cite{YCo5NP2006}, attributed to a Lifshitz transition when FB2 crosses $E_F$ \cite{FB-PRB2015,KoudelaPRB2015}. In Sm$_{\rm 2}$Co$_{\rm 17}$, the FB1 flat band near $E_F$ persists over a large $k_z$ range (Figure~S2 and Figure~S9 in SI). Such spin-polarized flat bands at $E_F$ open avenues for exploring emergent phenomena, e.g., high-temperature fractional quantum Hall effects \cite{FQHE1,FQHE2,FQHE3}.

The robust uniaxial FM in REPMs is often explained through $3d$-$3d$ and $3d$-$4f$ magnetic interactions. Our momentum-resolved measurements directly verify the itinerant (localized) character of the $3d$ ($4f$) electrons underlying these interactions, as illustrated schematically in Figure \ref{Fig5}(c): the $3d$-$3d$ coupling stems from itinerant Co $3d$ bands near $E_F$ and dominates the FM with a large band splitting of $\sim 1$\,eV (Figure~S6 in SI). In contrast, Sm $4f$ electrons remain mostly localized, and the exchange coupling $J$ of the $3d$-$4f$ interaction, estimated from DFT+$U$ calculations, is on the order of a few tens of meV---much smaller than the $3d$-$3d$ interaction. Furthermore, our DFT+$U$ results indicate that the $4f$ spin moments align parallel to the minority-spin direction of the $3d$ electrons (see SI), whereas the $4f$ orbital moments are antiparallel to (and larger than) their spin moments. Hence, the net moments of Sm $4f$ and Co $3d$ electrons are parallel and ferromagnetically coupled along the $c$-axis. These findings naturally account for the strong FM observed in Sm$_{\rm 2}$Co$_{\rm 17}$ \cite{SC5-Tmagmom-JAP1979}. 
Previous study has proposed that the $3d$-$4f$ interaction is most likely mediated by the small amount of Sm $5d$ states near $E_F$ \cite{campbell1972}, which is supported by our calculations in Figure~S6 in SI. In addition, because FB1 has a clear overlap with these $5d$ states, its nearly flat dispersion with large DOS can further reinforce the indirect $3d$-$4f$ exchange and thus help boost the uniaxial ferromagnetism.

The $3d$-$4f$ interaction is important for the large MCA in REPMs and is likely the cause of the anomalous temperature evolution of the Sm valence. Specifically, the enhanced Sm$^{3+}$ component at low temperatures in Sm$_{\rm 2}$Co$_{\rm 17}$ could be attributed to its strong FM order, which competes with the $4f$ delocalization process driven by Sm valence fluctuations, thereby suppressing itinerant (and nonmagnetic) Sm$^{2+}$ components. This scenario is analogous to the well-known competition between magnetic order and Kondo screening in Ce- and Yb- based Kondo systems \cite{Coleman2007heavy,gegenwart2008quantum,kirchner2020colloquium, zhaoSCPMA2025}. Our findings thus suggest that strong valence fluctuations (or the Kondo effect) could undermine the robust FM (and consequently the performance of REPMs), in agreement with recent theoretical studies on Ce-based REPMs \cite{CeCo5PRB2022,Galler2021}. Interestingly, the rise in the Sm$^{3+}$ component coincides with a pronounced increase in the coercive field $H_c$ at low temperature (Figure~\ref{Fig1}(f) and Figure~S1(d)). It would therefore be interesting to investigate how this valence change can be connected to the enhanced $H_c$ at low temperature \cite{3d-4f-book2007}.

\section{Conclusion}
In summary, we have successfully grown epitaxial Sm$_{\rm 2}$Co$_{\rm 17}$ films by MBE and revealed, for the first time, the the layer- and momentum-dependent electronic structure for a typical REPM. Our results unveil that the Co $3d$ electrons form flat bands near $E_F$, one at $\sim$\,--300\,meV and another right at $E_F$, which arise from the orbital-selective destructive interference in a kagome-honeycomb lattice and correlation effects, respectively. Our work further reveals that Sm$^{\rm 3+}$ $4f$ states are far away from $E_F$ and exhibit an anomalous temperature evolution, due to the $3d$-$4f$ interaction and competition with strong FM. These findings can be reasonably captured by theoretical calculations, thereby paving the way for a fundamental understanding of REPMs. The spectroscopic insight also hints on alternative approaches to search for high-performance REPMs, through judicious design/engineering of crystal structures and associated flat bands. Finally, our results also suggest new routes to explore emergent phenomena associated with flat bands near $E_F$ \cite{CaNi2nature2023,CuV2S4NP2024,NRM2024}, by generating destructive interference in kagome-honeycomb lattices in combination with strong correlation effects.

\textit{This work is supported by the National Key R$\&$D Program of China (Grant No. 2022YFA1402200, No. 2023YFA1406303, No. 2022YFA1403202), the National Science Foundation of China (No. 12174331, No. 12350710785), the Key R$\&$D Program of Zhejiang Province, China (2021C01002), the State Key project of Zhejiang Province (No. LZ22A040007), Zhejiang Provincial Natural Science Foundation of China (Grant No. LGG22E020006), Centro Nacional de Desenvolvimento Cient\'ifico e Tecnol\'ogico (CNPq) Grant No. 404312/2024-1 and and S\~{a}o Paulo Research Foundation (FAPESP) Garnt No. 2018/08845-3. We thank Prof. Frank Steglich, Prof. Zhentao Wang, Prof. Zheng Liu, Prof. Michael Smidman, Prof. Jiaying Jin, Prof. Yanwu Xie, Prof. Zhicheng Zhong, Prof. Xin Lu, Prof. Jiefeng Cao, Dr. Bodry Tegomo Chiogo, Dr. Ruiwen Xie, Dr. Shuaishuai Yin, Dr. Ge Ye and Ms. Xueshan Cao for discussions and help on experiments.}

\noindent \textbf{Conflict of interest }
The authors declare that they have no conflict of interest.

\noindent \textbf{Supporting Information }
The supporting information is available online at http://phys.scichina.com and https://link.springer.com. The supporting materials are published as submitted, without typesetting or editing. The responsibility for scientific accuracy and content remains entirely with the authors.

\bibliography{SmCoref}
\end{document}